# All-optical free-space routing of upconverted light by metasurfaces via nonlinear interferometry

*Agostino Di Francescantonio[1], Attilio Zilli[1], Davide Rocco[2], Laure Coudrat[3], Fabrizio Conti[1], Paolo Biagioni[1], Lamberto Duò[1], Aristide Lemaître[4], Costantino De Angelis[2], Giuseppe Leo[3], Marco Finazzi[1]\*, Michele Celebrano[1]*

[1]Physics Department, Politecnico Milano, Piazza Leonardo Da Vinci 32, 20133 Milano, Italy.
[2]Department of Information Engineering, University of Brescia, Via Branze 38, 25123 Brescia, Italy.
[3]Université de Paris, CNRS, Laboratoire Matériaux et Phénomènes Quantiques, 75013 Paris, France
[4]Centre de Nanosciences et de Nanotechnologies, CNRS, Université Paris-Saclay, 91120 Palaiseau, France

## Abstract

All-optical modulation yields the promise of high-speed information processing. In this frame, metasurfaces are rapidly gaining traction as ultrathin multifunctional platforms for light management. Among the featured functionalities, they enable light wavefront manipulation and, more recently, demonstrated the ability to perform light-by-light manipulation through nonlinear optical processes. Here, by employing a nonlinear periodic metasurface, we demonstrate all-optical routing of telecom photons upconverted to the visible range. This is achieved via the interference between two frequency-degenerate upconversion processes, namely third-harmonic and sum-frequency generation, stemming from the interaction of a pump pulse with its frequency-doubled replica. By tuning the relative phase and polarization between these two pump beams, and concurrently engineering the nonlinear emission of the individual elements of the metasurfaces (meta-atoms) along with its pitch, we route the upconverted signal among the diffraction orders of the metasurface with a modulation efficiency up to 90%. Thanks to the phase control and the ultrafast dynamics of the underlying nonlinear processes, free-space all-optical routing could be potentially performed at rates close to the employed optical frequencies divided by the quality factor of the optical resonances at play. Our approach adds a further twist to optical interferometry, which is a key-enabling technique in a wide range of applications, such as homodyne detection, radar interferometry, LiDAR technology, gravitational waves detection, and molecular photometry. In particular, the nonlinear character of light upconversion combined with phase sensitivity is extremely appealing for enhanced imaging and biosensing.

# Introduction

Optical interference is a physical phenomenon first reported by Thomas Young at the beginning of the 19th century, whereby two or more light waves combine producing spatial or temporal intensity fringes caused by the alternate addition and cancellation of the electric fields, ultimately ruled by their phase relation. Interference is the key-enabling mechanism at the heart of a wide range of applications, such as optical homodyne detection[1], radar interferometry,[2] LiDAR technologies,[3,4] sensing[5,6,7] (including quantum[8] sensing), enhanced imaging,[9,10,11] and molecular photometry.[12] Along with their effective application to sensing[13], Mach–Zehnder interferometers are extensively applied in integrated optics to realize fast (~100 GHz rates) modulation and routing of optical signals.[14,15,16] To this aim, all-optical switching, consisting in a light beam (signal) being modulated by a second beam (control), does not only grant ultrafast (~ 10 fs) operation but also extremely low energy consumption (~fJ) in integrated photonics platforms.[14,17,18,19]

Optical metasurfaces – ultrathin planar ensembles of nanostructures – can shape the wavefront of the propagating light by diffraction and refraction.[20,21] Thanks to the possibility of implementing multiple functionalities,[22] these platforms are soon expected to complement and even supplant bulk optics in many applications where a small footprint is desirable. Despite the latest achievements in meta-optics, to date the operation of metasurfaces beyond the linear and passive regime remains an open challenge. Nonlinear conversion[23,24,25,26] and active tuning/reconfigurability[27,28,29] of optical metasurfaces are two outstanding subjects that promise major advances. Stimulated and spontaneous parametric processes in metasurfaces[23,24,30,31] are indeed strategical for the development of infrared imaging,[32,33] THz generation[34] and detection,[35,36,37] and for the generation of entangled photon pairs.[38] Nonlinear metasurfaces were also suggested as efficient platforms for optical holography[39,40,41] and refractometric sensing,[42,43]. The striking development in nonlinear meta-optics is complemented by the recent breakthroughs in light steering and shaping with metasurfaces.[39,44,45] In particular, specific efforts have been recently devoted to speed up light modulation via metasurfaces and GHz rates were reported in the linear regime exploiting the electro-optic effect.[46]

Light-by-light modulation via metasurfaces would enable the realization of extremely compact and fast devices for optical communication in free space, which are virtually free from in-coupling and propagation losses. Recently, this has been realized in semiconductor platforms by modulating linear and nonlinear optical signals.[47,48,49] Although ultrafast dynamics down to the picosecond were achieved, no phase control has yet been reported, hence setting a limit to the achievable switching rates. Nonlinear parametric optical processes can be an effective mechanism to perform phase modulation. In this respect, Klimmer et al. recently reported the ultrafast and efficient all-optical modulation of the second-harmonic generation by an atomically thin semiconductor,[50] suggesting the possibility to employ phase control for the realization of high-speed frequency converters, nonlinear all-optical modulators, and transistors.

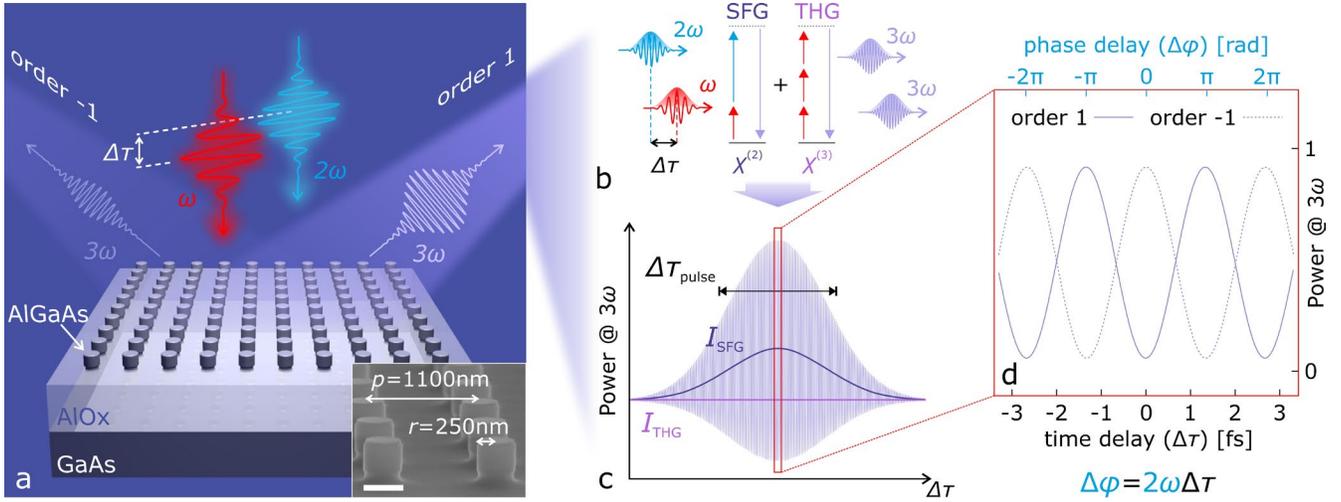

**Figure 1. Free-space routing of upconverted light by a dielectric metasurface.** a) Nonlinear upconversion of an ultrashort pulse ($\omega$) with its frequency-doubled replica ($2\omega$) mediated by an AlGaAs metasurface on AlOx substrate. The realized metasurface – *i.e.,* a periodic arrangement of nanocylinders – allows one to perform a directional sampling in the Fourier space, thereby breaking the detection symmetry. By engineering the metasurface periodicity, $p$, along with the emission by the individual meta-atom in the $k$ space, it is possible to route the light upconversion in the different diffraction orders of the metasurface. Switching between the orders can be attained by tuning the relative time/phase delay ($\Delta\tau/\Delta\varphi$) between the pump pulses. Inset: tilted-view scanning electron micrograph of the investigated metasurface. Scale bar: 500 nm. b) The adopted dual-beam pumping scheme ensures efficient generation of THG ($3\omega$) as well as SFG ($\omega + 2\omega$), which are frequency-degenerate. c) Interference between THG and SFG can occur in specific diffraction orders of the metasurface (*e.g.*, -1 and +1 in panel a) and, by varying the time delay $\Delta\tau$, an interference trace can be recorded. d) Opposite diffraction orders yield interference fringes that are in phase opposition. This implies that, by introducing a phase delay $\Delta\varphi = \pi$, the upconverted light at $3\omega$ can be routed between the orders. Given the employed wavelengths and that the beating angular frequency is $2\omega$ (see Ref. 51), this is equivalent to a pulse delay $\Delta\tau = \Delta\varphi/2\omega = \pi/2\omega \approx 1.3$ fs. By optimizing pump powers, periodicity and nanocylinder geometry, a routing efficiency (*i.e.,* fringe visibility) of almost 90% is achieved.

In this work, we exploit the interference between two upconversion pathways that shift telecom photons ($\lambda$ = 1551 nm) to visible wavelengths ($\lambda$ = 517 nm) to perform all-optical routing by means of an Aluminum Gallium Arsenide (AlGaAs) periodic nonlinear metasurface (see Figure 1a). Typical upconversion mechanisms involve either the interaction of energy-degenerate photons – *e.g.* second- and third-harmonic generation (SHG and THG) – or photons of different energies – *e.g.* difference- and sum-frequency generation (DFG and SFG). Here, we perform upconversion via both THG and SFG through a dual-beam pumping scheme (see Figure 1b), where an ultrashort telecom pulse ($\omega$) combines with its frequency-doubled replica ($2\omega$). Although in this configuration THG and SFG are frequency-degenerate (at $3\omega$), by symmetry arguments the interference averages to zero in the far-field for normal-incidence illumination in an individual AlGaAs nanocylinder.[51] This stems from the even (odd) number of pump photons involved in SFG (THG), which makes SFG (THG) even (odd) with respect to the illumination axis in all geometries where the latter coincides with a $C_2$ symmetry axis. Here, by engineering the pitch of an AlGaAs nonlinear metasurface, we were able to perform a directional sampling of the upconverted light from individual meta-atoms in the $k$-space. This allows to break the detection symmetry in the

Fourier space, hence enabling the constructive/destructive interference between the two processes in the diffraction orders of the metasurface (see Figure 1c). Employing the relative time/phase delay between the pump pulses as a tuning knob, we achieve all-optical routing of the light upconverted by the metasurface between different sets of diffraction orders (see Figure 1d). By optimizing the pump powers, metasurface periodicity and nanocylinder geometry we demonstrate a modulation amplitude up to 90%. The polarization state of the pump and emitted beams is also employed to reconfigure the routing of the upconverted light between different sets of diffraction orders. This, along with the accurate phase control we demonstrate, sets a significant milestone towards ultrafast and reconfigurable all-optical logic operation with meta-optics.

**Interferometric routing with a nonlinear metasurface**

The details of the experimental setup are described in the Methods section (Figure S1c). Briefly, ultrashort pulses ($\Delta\tau_p \sim 160$ fs) at telecom wavelengths ($\omega$) are combined with their frequency-doubled replica ($2\omega$) on the metasurface. Both beams are focused on the back focal plane (BFP) of the objective to realize a quasi-collimated illumination with illumination diameters of about 25 μm (corresponding to about 400 illuminated nano-pillars). The upconverted light generated at $3\omega$ via both SFG and THG by the metasurface is collected with the same focusing objective (numerical aperture NA = 0.85) in an epi-detection reflection scheme and separated from the excitation path by a long-pass dichroic mirror (band-edge 650 nm). By means of an additional lens ($f$ = 500 mm) in the collection path, we image the BFP of the objective into a cooled CCD camera. A delay stage (resolution $\tau_{res}$ = 0.66 fs) is inserted in the path of the $\omega$ beam for a coarse adjustment of the time delay between the pulses, while an additional compensated half-wave liquid-crystal retarder (LCR) provides a finer adjustment of the phase delay between the pulses.

The metasurface is composed of AlGaAs nanocylinders (the meta-atoms) on an AlOx/GaAs substrate (see Fig. 1a).[52] The details on the sample fabrication are presented in the Methods Section. The nanocylinders have height $h$ = 400 nm and radius $r$ = 250 nm. For this geometry, a field distribution corresponding to an electric dipole is excited by the $\omega$ ($\lambda$ = 1551 nm) beam within each nano-pillar. The metasurface pitch, $p$, is varied from 900 nm to 1500 nm to optimize the diffraction orders with respect to the nonlinear emission distribution (form factor) of the single nano-pillar. The inset of Figure 1b is an electron microscope image of the metasurface featuring $p$ = 1100 nm, which provides first diffraction orders at $NA = \lambda_{THG/SFG}/p \sim 0.47$, corresponding to ~28° (see Figure 2a).

Figure 2a shows a typical BFP image of the upconverted light by the metasurface with $p$ = 1100 nm obtained by averaging 40 BFP frames acquired while varying the pulse delay around zero in steps of $\tau \approx 0.66$ fs. It can be noticed that the average upconverted power is evenly distributed between the first diffraction orders. Conversely, individual frames acquired at a specific phase delays between the pulses show a strong power modulation between opposite diffraction orders. Figure 2b shows the difference between two BFP frames acquired at a relative delay $\Delta\tau \approx 1.3$ fs between the pump pulses, corresponding to a phase shift $\Delta\varphi = \pi$ for the optical field oscillation at the beating frequency $2\omega$ (see below)[53]. We arbitrarily set $\Delta\varphi = 0$ around zero delay (i.e. pump pulse temporal superposition) when the intensity of the (0,+1) diffraction order is maximized (see also Figure 1d). The interference traces for each diffraction order are retrieved by integrating the intensity emitted over each diffraction spot in the BFP maps as a function of $\Delta\tau$ (Fig. 2c–e). The traces show a Gaussian envelope with FWHM $\Delta\tau_{\text{meas}} = 240$ fs, which agrees with the estimated FWHM $\Delta\tau_{\text{est}} = \Delta\tau_{\text{p}} \cdot 1.41 \cong 225$ fs of the convolution envelope of the $\omega$ and $2\omega$ pulses.

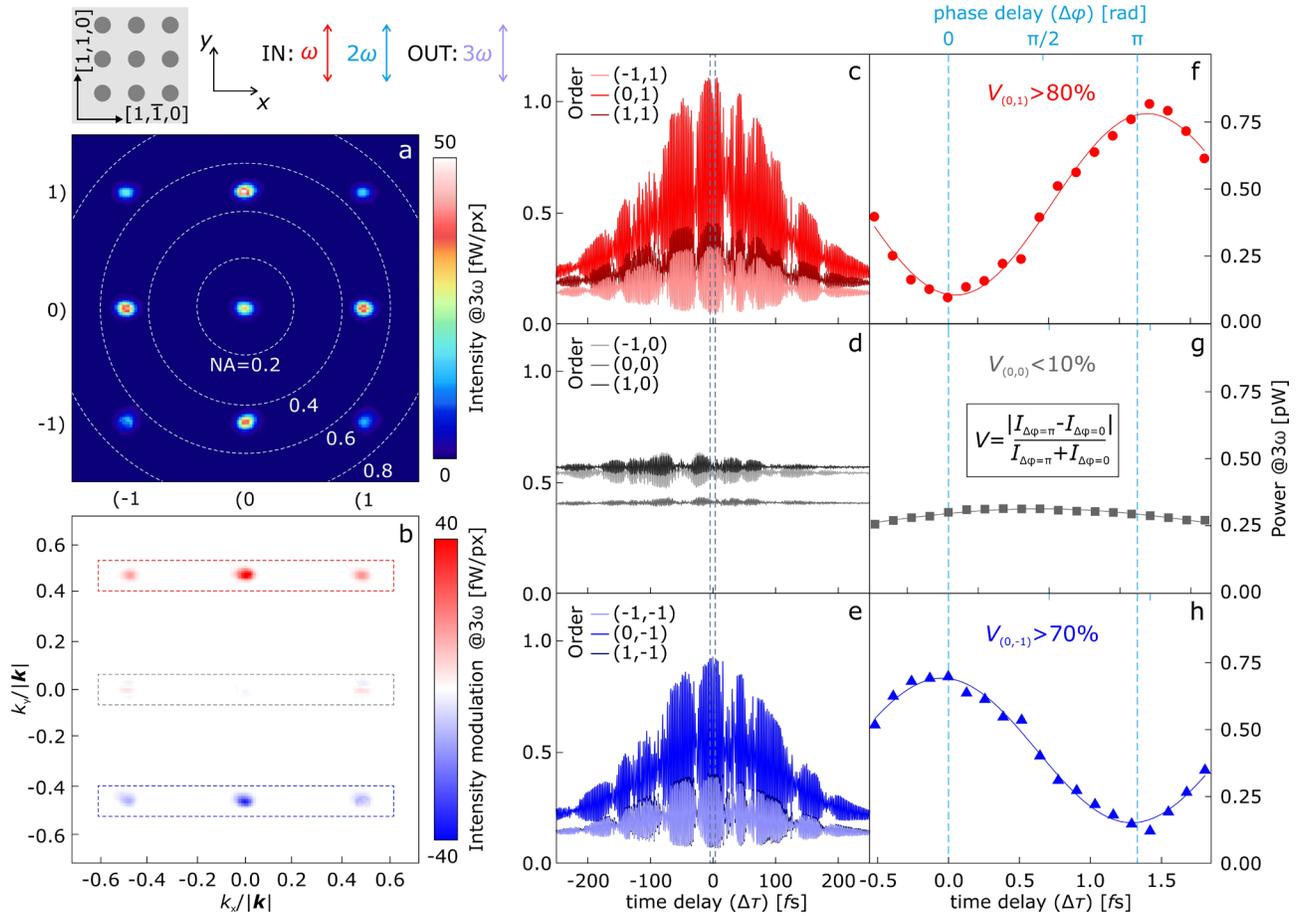

**Figure 2. Upconversion and routing with a nonlinear metasurface.** a) Back focal plane (BFP) image of upconverted intensity (*i.e.* optical power per pixel) at $3\omega$ (SFG+THG) by an AlGaAs metasurface with lattice periodicity, $p = 1100$ nm. The image is obtained by averaging 40 frames acquired at 0.66 fs relative delay steps between the pump pulses (about 27 fs delay span), while the intensity in fW/pixel is estimated from the counts by accounting for the collection throughput (see Materials and Methods). The pump powers are $P_\omega \approx 17$ mW and $P_{2\omega} \approx 11$ μW. The linear polarizations of both pumps (IN) and emission (OUT) are set parallel to the [1,1,0] crystal axis (parallel to the $y$ axis, see sketch above panel a). The white dashed circles identify the numerical aperture (NA = $n \sin\theta = k_\parallel/|\mathbf{k}|$) scale. Diffraction orders (0,1), (0,−1), (1,0), (−1,0) correspond to NA = 0.47 ($\theta = 28°$). b) An image obtained by subtracting two BFP images acquired at a relative delay of 1.33 fs. Panels c, d and e show the upconversion modulation delay traces caused by the interference between SFG and THG retrieved from the diffraction orders in maps a, b. The traces are sampled every $\tau = 0.66$ fs, which is the maximum resolution of the delay stage. The power from each diffraction spot is obtained by integrating the intensity collected in a 13×13-pixel area around the centroid of the diffraction order and quantified by accounting for the optical losses of the setup and detector photon-to-electron conversion efficiency at $3\omega$ (see the Methods section). Panels f, g and h show expansions of the (0,X) orders traces around $\Delta\tau = 0$ (dashed area in c, d and e). The dots represent the experimental data, while the solid lines are a sinusoidal fit to the data. The light blue dashed lines indicate the phase delays corresponding to the frames subtracted to obtain panel b. The (0,±1) orders show highest modulation, reaching a visibility, $V$, larger than 80% (definition in panel g).

The upconverted signal at the $(0, \pm 1)$ diffraction orders is efficiently modulated as a function of the pump pulse delay. Conversely, the $(\pm 1, 1)$ and $(\pm 1, 1)$ orders show a weak modulation and the $(\pm 1, 0)$ and $(0,0)$ orders a negligible one. Thus, this approach allows the efficient routing of the upconverted light between the diffraction orders. By adjusting the power ratio between the pumps to $P_{2\omega}/P_\omega \approx 7 \times 10^{-4}$, we arrive to a maximum visibility of the fringes $V \sim 90\%$. To equalize the upconverted power beamed into different points in the $k$-space and gain full control over this platform, it is crucial to finely adjust the phase delay. Therefore, we inserted the LCR in the optical path of the $\omega$ beam with its slow axis parallel to the linear polarization of the beam. By varying the applied voltage between 0 and 10 V, a relative phase delay $\Delta\varphi$ between 0 and $2\pi$ (*i.e.* half wavelength at 1551 nm), which corresponds to a full interference period (*i.e.* about 2.67 fs). In this way we could achieve a resolution of about $\pi/10$ in phase, which corresponds to a delay of about 150 attoseconds. Figures 2f–h display the upconverted power by the $(0, X)$ orders obtained by setting the delay stage around $\Delta\tau = 0$ and finely tuning the phase delay by means of the LCR (dashed grey areas in panels c–e). The high phase resolution allows to highlight that the $(0, \pm 1)$ orders are in precise phase opposition. This demonstrates that the metasurface allows performing efficient all-optical routing of upconversion between this set of orders by delaying the pump pulses of about $\Delta\tau_{\text{route}} = 1.3$ fs – *i.e.*, by introducing a $\Delta\varphi = \pi$ phase delay between the pump beams.

## Routing mechanism and efficiency optimization

To understand the upconversion routing mechanism operated by our metasurface one needs to resort to the emission of upconverted light from an individual nanocylinder (as recently discussed in Ref. 51), which provides the form factor modulating the diffraction spot intensities. Upon excitation with a $\omega+2\omega$ dual-beam pump configuration, the far-field projections of THG and SFG emitted power by an individual AlGaAs nanocylinder can significantly differ, since they depend on both the Mie-type resonances and the selection rules at play in the two separate processes. Being degenerate in wavelength, the fields of the two processes coherently superimpose in the far-field producing a well-defined intensity distribution in the Fourier space. Figures 3a and b show the far-field intensities at $3\omega$ computed with a Finite- Element Methods (FEM) simulation (COMSOL Multiphysics) assuming a dual-beam plane-wave excitation with intensities corresponding to the powers employed in our experiment. The RETOP toolbox[54] was employed to attain the exact far-field projection in a inhomogeneous space due to the presence of the substrate (see the Methods Section). The two maps show the BFP-projected intensity at $3\omega$ for $\Delta\varphi = 0$ and $\pi$, respectively. In this work, we selected nanocylinders with a radius of 250 nm, which support an electric-dipole resonance at $\omega$. This results in a simple form factor for the intensity with a single marked lobe around $k_x = 0$ and $k_y > 0$ ($k_y < 0$) for $\Delta\varphi = 0$ ($\Delta\varphi = \pi$). However, we stress that the form factor can be changed by exploiting different resonances, hence providing high flexibility to this routing approach. The power emitted as a function of time delay by the individual nanocylinder at $3\omega$ – *i.e.,* intensity integrated over the whole collection angles – can be modeled as[53]

$$P_{3\omega}(\tau) = P_{\text{THG}} + P_{\text{SFG}} \left(2^{-\frac{16\Delta\tau^2}{3\sigma^2}}\right) + 2\gamma\sqrt{P_{\text{THG}} \cdot P_{\text{SFG}}} \left(2^{-\frac{16\Delta\tau^2}{3\sigma^2}}\right) \cos(\Delta\varphi + \Delta\varphi_0), \quad (1)$$

where the modulation amplitude is proportional to the integral coefficient

$$\gamma = \frac{\int_{\Omega_{\text{det}}} \mathbf{E}_{\text{THG}}(k_x, k_y) \cdot \mathbf{E}_{\text{SFG}}(k_x, k_y) \, d\Omega}{\sqrt{\int_{\Omega_{\text{det}}} |\mathbf{E}_{\text{THG}}(k_x, k_y)|^2 \, d\Omega \int_{\Omega_{\text{det}}} |\mathbf{E}_{\text{SFG}}(k_x, k_y)|^2 \, d\Omega}} \quad (2)$$

that quantifies the directional and polarization overlap of the THG and SFG far-field-projected fields. In these expressions, $\mathbf{E}_{\text{SFG}}(k_x, k_y)$ and $\mathbf{E}_{\text{THG}}(k_x, k_y)$ (from here on $\mathbf{E}_{\text{SFG}}$ and $\mathbf{E}_{\text{THG}}$ for simplicity) are the (real) field amplitudes multiplied by the polarization unit vectors of the outcoupled SFG and THG fields in the direction defined by the $k_x$ and $k_y$ projections of the wavevector **k** onto the BFP. The phase difference $\Delta\varphi$ is defined as $\Delta\varphi = 2\omega\Delta\tau$, $\Delta\tau$ being the time delay between the $\omega$ and $2\omega$ pulses and $\Delta\varphi_0$ is the phase difference between $\mathbf{E}_{\text{SFG}}$ and $\mathbf{E}_{\text{THG}}$ at $\Delta\tau = 0$ (in Figs. 2f–h, $\Delta\varphi_0$ has been set, somewhat arbitrarily, equal to zero). A perfect superposition of the fields is realized for $P_{\text{THG}} = P_{\text{SFG}}$ and $\gamma = 1$, a condition that is obtained when $\mathbf{E}_{\text{THG}} \parallel \mathbf{E}_{\text{SFG}}$ and leads to perfect cancellation/summation depending on the phase delay between the $\omega$ and $2\omega$ pulses. However, in systems possessing a $C_2$ symmetry, such as a nanocylinder or a bulk material, $\gamma = 0$.[51] This is evident here since, while the intensity is directionally modulated between Figure 3a and 3b (see Figure 3c), the collected power over the full NA is not. This holds in general for any phase difference as a consequence of the opposite parity of the THG and SFG processes with respect to the optical axis of collection. Yet, one can retrieve interference by breaking the detection symmetry, for instance by collecting the $3\omega$ signal *in specific locations* in the Fourier space. This is done here by the designed diffracting metasurface of nanocylinders. In this way, the $3\omega$ intensity in the diffraction orders will become a function of the phase difference between the $\omega$ and $2\omega$ beams, enabling routing the upconverted light in different directions. This process can be optimized by selecting the geometry of the meta-atoms and their arrangement to have the maxima of the function $\mathbf{E}_{\text{THG}} \cdot \mathbf{E}_{\text{SFG}}$ modulating the form factor in Eq. [1] aligned with the $(k_x, k_y)$ direction of one of the diffraction orders of the metasurface. To this aim, we have computed the $\mathbf{E}_{\text{THG}} \cdot \mathbf{E}_{\text{SFG}}$ modulation form factor by performing numerical simulations, applying Floquet periodic conditions to the elementary cell of the individual pillar to account for possible near-field coupling between the meta-atoms (see Methods section). We have then selected the pitch value optimizing $\mathbf{E}_{\text{THG}} \cdot \mathbf{E}_{\text{SFG}}$ and aligning the maxima of the predicted visibility pattern (Figure 3d) with the $(0, \pm 1)$ diffraction orders. In this way, we maximize the modulation amplitude of the interference fringes, i.e., their visibility ($V$), defined as $V = (I_{\Delta\varphi=0} - I_{\Delta\varphi=\pi})/(I_{\Delta\varphi=0} + I_{\Delta\varphi=\pi})$. Using field intensities for the two pump plane waves that reproduce the powers employed in the experiment and computing the visibility $V$ for the $(0, \pm 1)$ orders, we retrieved the plot reported in Figure 3e (grey dots). This confirms that the metasurface showing best visibility in these experimental conditions is the one with pitch $p = 1100$ nm. This is also corroborated by experimental measurements performed on different metasurfaces with different pitches (see blue and red dots in Figure 3e). The deviation for smaller pitches is attributed to possible discrepancies due e.g. to a non-perfect collimation or tilting of the beam in the experiment and to slight deviations of the radius and pitch size of the realized metasurface with respect to the nominal values. We again stress that the possibility to change the nanocylinder size (*i.e.* main resonance at play) to feature a different resonating behavior of the meta-atoms and, consequently, a different $\mathbf{E}_{\text{THG}} \cdot \mathbf{E}_{\text{SFG}}$ distribution, makes this approach extremely versatile.

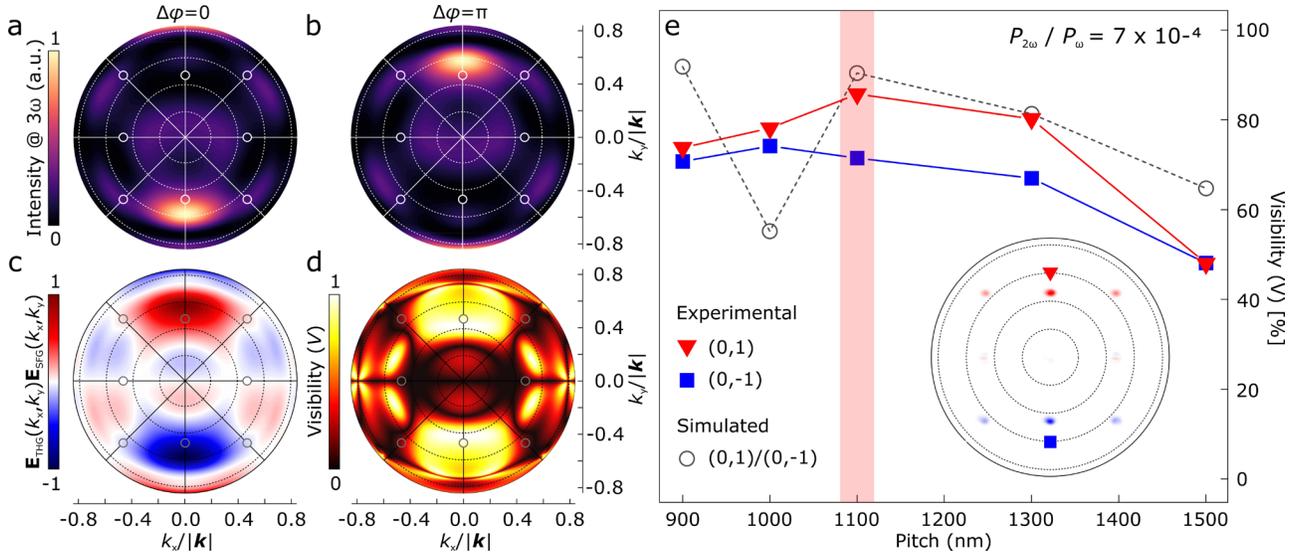

**Figure 3. Interferometric routing numerical optimization.** a, b) Simulated total upconversion intensity at $3\omega$ by an individual AlGaAs nanocylinder computed with COMSOL Multiphysics combined with RETOP far-field projection. Simulations are run assuming a power ratio $P_{2\omega}/P_\omega = 7 \times 10^{-4}$ and a co-polarized excitation and detection along $y$. While the overall intensity remains constant by introducing a phase shift $\Delta\varphi = \pi$ between the pulses, due to the constructive/destructive interference between SFG and THG the intensity profile is reshaped from a lobe at $k_y > 0$ and $k_x \simeq 0$ to $k_y < 0$ and $k_x \simeq 0$. Zero phase delay is arbitrarily set where the (0,1) diffraction order is maximized. c) Intensity modulation amplitude in the Fourier plane, provided by $\mathbf{E}_{\text{THG}}(k_x, k_y) \cdot \mathbf{E}_{\text{SFG}}(k_x, k_y)$, see Eq. (2). d) Simulated visibility in the Fourier plane calculated as $V = (I_0 - I_\pi)/(I_0 + I_\pi)$. The black/white circles in panels a–d show the location of the diffraction orders of the $p = 1100$ nm metasurface in the Fourier plane. e) Simulated visibility of the interference trace as a function of the pitch, $p$ for the (0,±1) orders (grey dots/dashed line), compared to the experimental visibility of the (0,+1) (red) and (0, −1) (blue) orders (see also inset) acquired at a pump power ratio $P_{2\omega}/P_\omega = 11~\mu\text{W}/17~\text{mW} \sim 7 \times 10^{-4}$ and co-polarized polarizations. The red bar marks the pitch $p$ of the metasurface presented in Fig. 2, which exhibits the highest experimental routing efficiency at this power ratio.

## Polarization-based reconfigurable routing

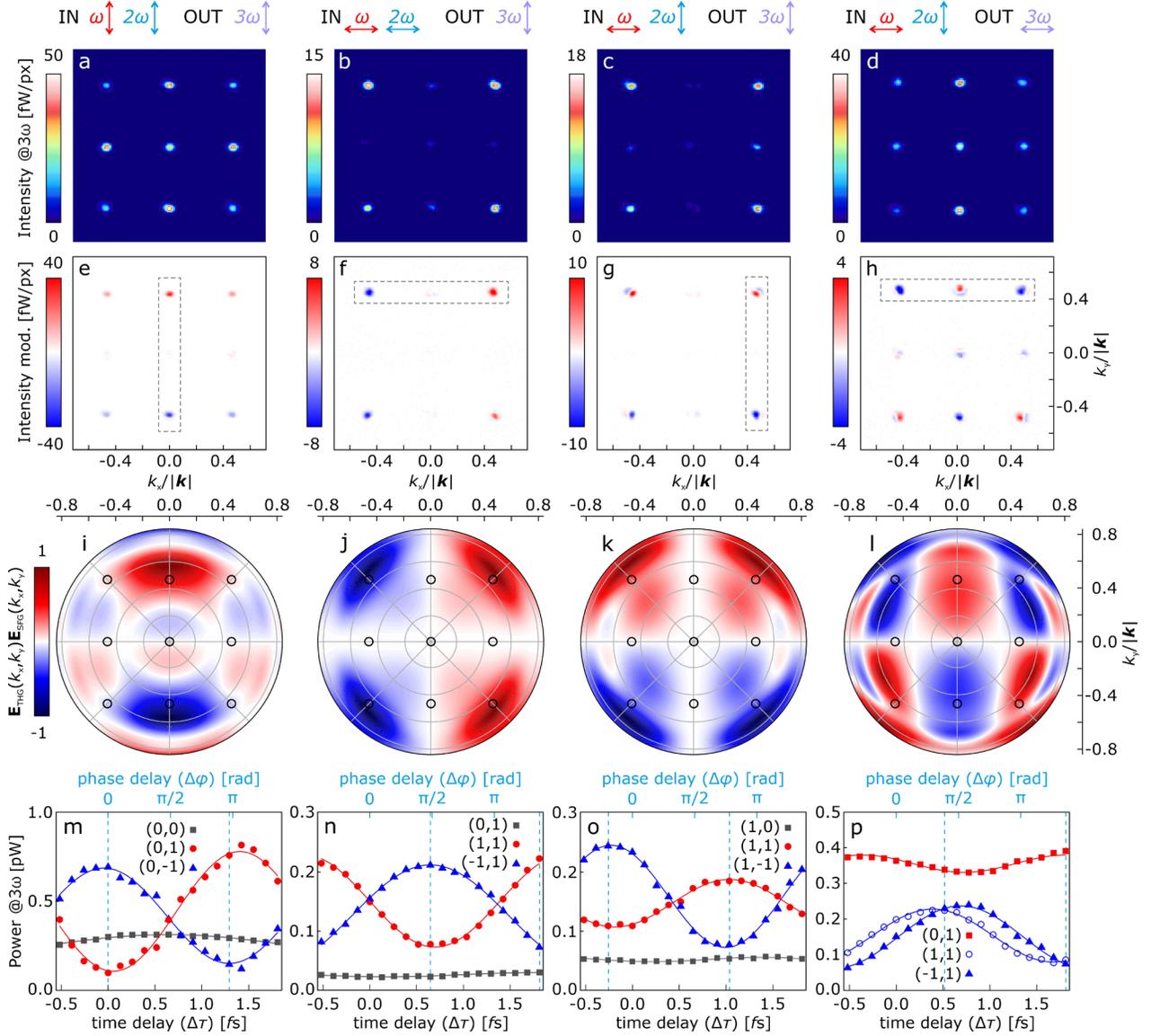

**Figure 4. Polarization-controlled routing** a–d) BFP images of the nonlinear emission at $3\omega$ by an AlGaAs metasurface with lattice periodicity $p = 1100$ nm, for different combinations of the input and output linear polarizations (red, light blue and violet arrows displayed at the top of each column). The images are obtained by averaging 40 frames acquired at 0.66 fs steps between the pump pulses around the zero-delay condition (about 27 fs delay span) as in Figure 2. e–h) Intensity modulation obtained by following the same procedure as for Figure 2. i–l) FEM simulations of $\mathbf{E}_{\text{THG}}(k_x, k_y) \cdot \mathbf{E}_{\text{SFG}}(k_x, k_y)$ computed as in Figure 3 for each polarization configuration. The small black circles in each map indicate the nominal position of the diffraction orders. m–p) Delay traces of the power collected at the main diffraction orders involved (symbols). The power of each diffraction is obtained by integrating the intensity acquired in a 13×13 pixels area around the centroid of the diffraction spot in the BFP maps at each delay, as in Figure 1. The solid lines are sinusoidal fits to the experimental data, while the dashed light blue lines represent the phase delays corresponding to the frames subtracted to obtain the images in panels e-h.

The versatility of our platform is augmented by the possibility to commute the all-optical routing among different sets of diffraction orders using the beams polarization as a selection switch. This is effectively equivalent to a reconfigurable routing between various combinations of $k_x$ and $k_y$ coordinates in the Fourier space. Figures 4a–d show the BFP images of the metasurface with $p = 1100$ nm reported in Figure 2 employing four different configurations of the polarizations of the $\omega$ and $2\omega$ pumps and of the emitted light at $3\omega$. It is straightforward to note that depending on the polarization configuration different sets of diffraction orders emit more efficiently. This is the result of the polarization dependence of the upconverted light form factor by the individual nanocylinder. In particular, while panels a and d display larger powers for the zero-diffraction order and the $(\pm 1, 0)$ and $(0, \pm 1)$, panels b and c maximize the $(\pm 1, \pm 1)$ orders. We stress that the analyzer placed in the detection path to select the linear polarization at $3\omega$ decreases the total detected power but allows increasing the modulation contrast and selecting the far-field region of interest. Figures 4e–h display the intensity modulation for the same polarization sets as in a–d, obtained by subtracting two BFP images acquired at a relative delay $\Delta \tau_{\text{route}} \approx 1.3$ fs between the pulses and corresponding to a $\Delta \varphi = \pi$ dephasing at the $2\omega$ carrier frequency. The simulated form factors $\mathbf{E}_{\text{THG}} \cdot \mathbf{E}_{\text{SFG}}$ underpinning the modulation amplitude (as in Figure 3c) are reported in Fig. 4i–l for different combinations of the polarizations of the $\omega$, $2\omega$, and $3\omega$ beams. They all show an antisymmetric character with respect to $\pi$ rotations around the origin of the Fourier space, which stems from the aforementioned parity properties $\mathbf{E}_{\text{THG}}(k_x, k_y) = \mathbf{E}_{\text{THG}}(-k_x, -k_y)$ and $\mathbf{E}_{\text{SFG}}(k_x, k_y) = -\mathbf{E}_{\text{SFG}}(-k_x, -k_y)$. Moreover, the selected combinations of polarizations have a well-defined parity also with respect to reflections across the $xz$ and $yz$ planes. This results in $\mathbf{E}_{\text{THG}} \cdot \mathbf{E}_{\text{SFG}}$ being *even* (*odd*) with respect to a reflection across the plane *parallel* (*perpendicular*) to the electric field of the $2\omega$ beam. Finally, Figures 4m–p shows the time(phase) traces acquired over a whole $\pi$ shift dephasing between the pump beams corresponding to the diffraction orders selected in panels e-h (see dashed grey areas). Note that the phase difference at zero delay between the routing orders deviates from 0 (or $\pi$) because of a nonzero phase difference $\Delta \varphi_0$ between $\mathbf{E}_{\text{SFG}}$ and $\mathbf{E}_{\text{THG}}$, see Eq. (1). We ascribe this to the static phase delay intrinsically introduced by the half waveplates employed to rotate the beams polarization and to long-term mechanical drifts in the interferometer.

One of the key features of the proposed approach is that the upconverted radiation is efficiently routed among the diffraction orders featuring higher powers, while weaker orders mostly show negligible modulation. This mechanism can be verified by comparing the averaged images in panels a–d to the differential ones in panels e–h, and is supported by the time/phase traces in panels m–p. This emphasizes that most of the upconverted power is processed by the metasurface, pointing towards the all-optical reconfigurability of routing among different sets of diffraction orders. For example, while in panels a,e the modulation amplitude is larger for the $(0, +1)$ and $(0, -1)$ diffraction orders, in panels b,f and c,g routing occurs between the $(-1, \pm 1)$ and $(+1, \pm 1)$ or $(\pm 1, +1)$ and $(\pm 1, -1)$, respectively. Therefore, while a co-polarized state (panel e) promotes routing mainly between the two diffraction orders at $k_x = 0$, in the other two configurations (panels f,g), the modulated diffraction orders can be selectively addressed by changing the polarization state of one of the two pump beams. In panels d,h routing becomes even richer, although this comes at the expense of routing selectivity. In fact, by increasing the number of the $\mathbf{E}_{THG} \cdot \mathbf{E}_{SFG}$ nodes, the optimal conditions for visibility are hardly met for all orders. Indeed, the THG and SFG intensity distributions by an individual pillar show a non-trivial sensitivity to the polarization state of the input/output beams, and on the electromagnetic modes excited inside the nano-pillars.[51] This is why we specifically opted for a nanocylinder geometry that allows to obtain the simplest upconversion intensity distribution in the Fourier plane (*i.e.* form factor), hence, maximizing the routing efficiency. Panels i–l show the simulated intensity modulation obtained by FEM (similar to Figure 3c), displaying an excellent agreement with the experimental power modulation maps for the regions identified by the circles (*i.e.* the nominal position of the diffraction orders).

These results highlight the robustness and reliability of this approach and allow envisioning its extension to other materials and designs. Indeed, following the same approach, we also realized all-optical routing using a metasurface with a diamond lattice arrangement of the meta-atoms (namely, with nearest neighbors along the [100] and [010] crystal axes) instead of the square one presented here, showing that all-optical routing between other direction sets can be easily accessed. In addition, it is also possible to span the Fourier plane by employing a metasurface with larger/narrower pitch as well as tuning the modulation form factor by engineering the meta-atom sizes and shapes. We stress that, although in this proof of concept we limited ourselves to static measurements, dynamic routing using phase modulation in this platform can potentially reach speeds higher than the THz, provided that enough upconversion intensity is achieved. Indeed, the reconfiguration time of the emission can be up to $\Delta\tau_{route} \cdot Q \approx 30$ fs, where $Q\sim20$ represents the quality factor of the resonant modes at play. Finally, we emphasize that the possibility to monitor two or more parallel routing channels inherently enables a differential signal read-out. This can be extremely beneficial for applications such as nonlinear sensing,[42,43] where the capability to extract a differential signal allows one to rule out possible intensity fluctuations, hence enhancing the sensitivity.

## Conclusions

We have demonstrated the upconversion and routing of telecom photons to the visible range via periodically engineered AlGaAs metasurfaces. Upconversion is achieved simultaneously through THG and SFG, which are wavelength-degenerate due to adopted $\omega$+2$\omega$ dual-beam pumping scheme, where $\omega$ falls within the third telecommunication window. By exploiting the interference between the two processes and tuning the relative phase between the pump pulses with a resolution $\pi/10$ (*i.e.* about 150 attoseconds time delay) we attained all-optical routing among different metasurface diffraction orders with a modulation amplitude up to 90%. We could reach an upconversion figure of merit of telecom photons to the visible $\beta = P^{3\omega}/P^{2\omega}P^{\omega} > 10^{-5}W^{-1}$ using pump fluences below $1.2 J/m^2$. We also demonstrate that the polarization of both input and output beams can be employed as a further degree of freedom to reconfigure the routing process between different sets of diffraction orders. By employing high-$Q$ platforms[55] providing larger conversion efficiencies, the proposed approach may offer a compact method to all-optically demultiplex in free space telecom signals between various detection channels in the visible range at very high speed. Concurrently, since the two pump frequencies experience different refractive indexes in any dispersive medium, their phase relation and, hence the upconverted light, can be employed as a sensitive refractometric probe. In addition, as already mentioned, the presence of multiple output channels is ideal for differential measurements, which allow to compensate for intensity instabilities. All these features, along with the nonlinear character of the underlying processes, is extremely appealing for sensing applications.[42,43] In this frame, it has been recently suggested that the interference between THG and SFG in metasurfaces [56] can be effectively employed for chiral sensing. Finally, the extreme phase sensitivity of the upconverted light in these platforms can be also thought for future LiDAR applications where compact telecom ultrafast laser sources are employed.[57]

**Materials and Methods**

Fabrication

Metasurfaces were fabricated in a similar way as described in ref. [1], starting from a 1.5 µm-thick $Al_{0.98}Ga_{0.02}As$ film and a $h$=400 nm-thick layer of $Al_{0.18}Ga_{0.82}As$ successively grown on a GaAs (001) substrate by molecular-beam epitaxy. The nanostructures are lithographically patterned out of the $Al_{0.18}Ga_{0.82}As$ layer, with nanocylinders target radius $r$ of 250 nm. A final step of selective oxidation of the Al-rich film creates a low-refractive-index ($n$ = 1.6) AlOx layer, which favours an effective field confinement in the meta-atoms ($n$=3.2). To explore different diffraction patterns, a set of metasurfaces with both square and diamond lattice arrangement with respect to the in-plane ⟨110⟩ crystallographic axes was fabricated. The lattice pitch $p$ is varied from 900 nm to 1500 nm. Representative scanning electron microscope (SEM) images of the metasurfaces with the AlGaAs crystallographic axes are shown in Figure S1a,b.

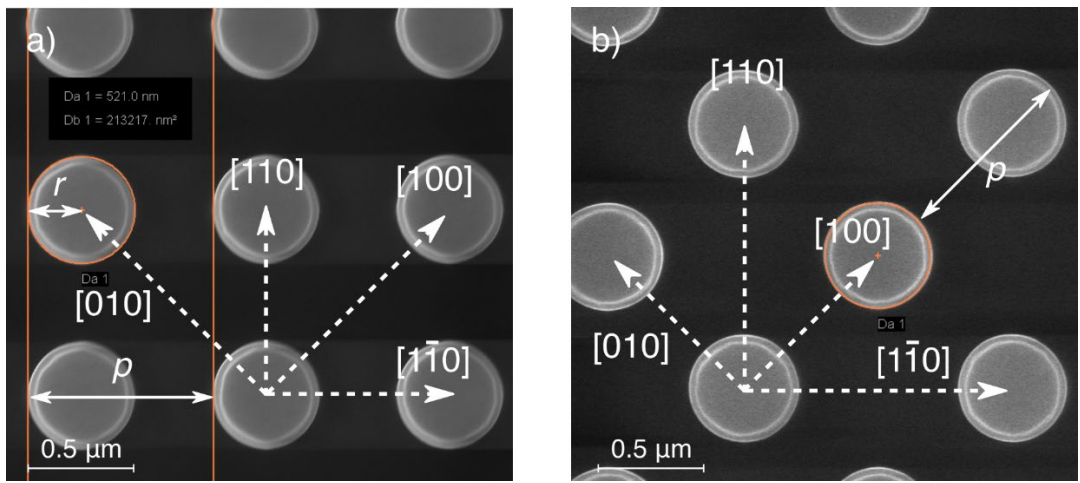

**Figure S1**. a,b) SEM micrographs of the metasurfaces in (a) square and (b) diamond arrangement, with $r$ = 250 nm and $p$ = 900 nm.

Experimental Setup

The linear optical properties of the metasurfaces are characterized by spectral analysis of the radiation reflected by the sample. The broadband light from a stabilized fibre-coupled white-light source with black-body spectrum in the wavelength range from 360 nm to 2600 nm (Thorlabs, SLS201/M) is focused on the sample by a 0.25 NA objective (Olympus RMS10×, 10×, 0.25 NA, Plan Achromat Thorlabs Inc.). The reflected radiation is collected by the same objective and separated from the incoming beam by means of a polarization-conserving beam splitter. The collected radiation is spatially filtered by an iris placed in an intermediate image plane. The filtered signal is then focused onto the input slit of a spectrometer (Andor, ShamrockSR-303i) equipped with a 150 grooves/mm ruled diffraction grating and a back-illuminated charge-coupled device (CCD) camera (Andor, iKon-M DU934P-BV).

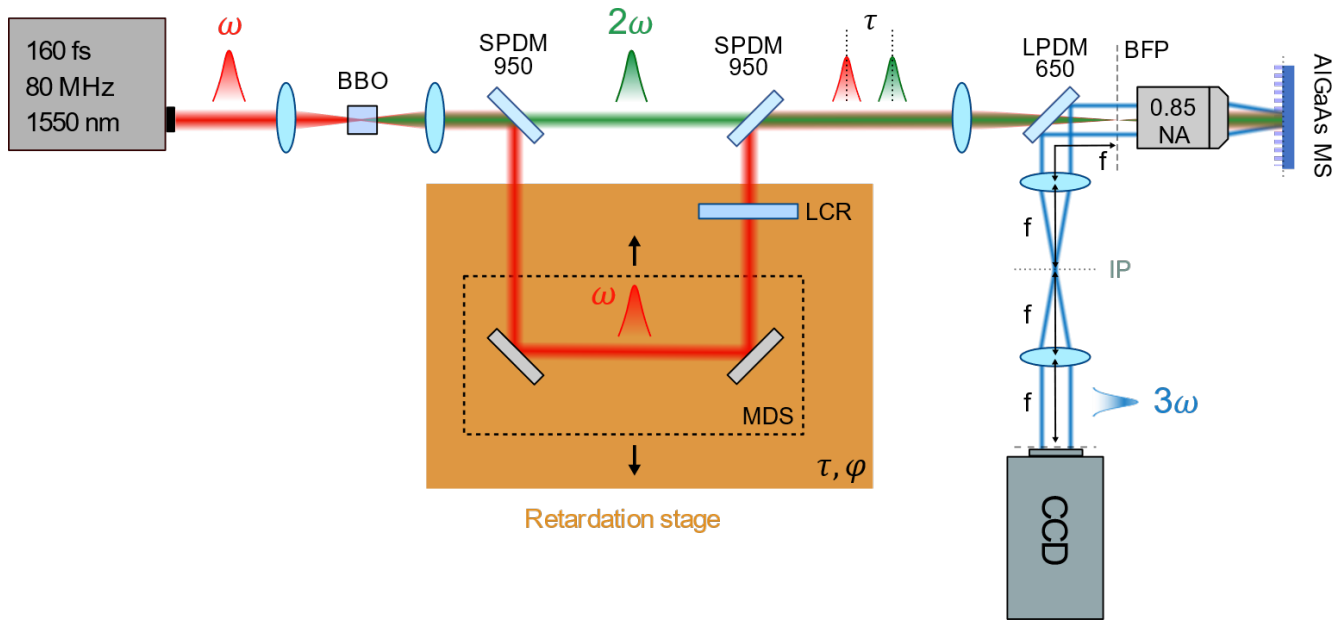

**Figure S2**. Layout of the experimental set-up. List of acronyms: BBO = β-barium- borate, SPDM = short-pass dichroic mirror, MDS= mechanical delay stage, LCR = liquid-crystal retarder, LPDM = long-pass dichroic mirror, BFP = back focal plane, NA = numerical aperture, MS = metasurface, IP = image plane.

The two-pump experiment (see Figure S2 and ref. [2]) was performed by employing a soliton mode-locked Er:Yb:glass laser (OneFive, Origami 15-80), which provides the beam at the fundamental wavelength (FW) 1550 nm (angular frequency $\omega$), with 160 fs pulse duration and 80 MHz repetition rate. The FW is partially duplicated in frequency via second-harmonic generation (SHG) by a β-barium borate (BBO) crystal (Eksma optics, BBO-SHG@1554nm), resulting in a beam with a wavelength of 775 nm (angular frequency $2\omega$). The paths of the two pulses are then separated by means of a short-pass dichroic mirror (Thorlabs, DMSP950). The $\omega$ beam path length is adjusted via a linear mechanical delay stage (MDS) (Physik Instrumente, M-404), delivering a minimum delay step of about 0.66 fs. In addition, a liquid-crystal variable retarder (LCR) (Thorlabs, LCC1411-C) is placed in the $\omega$ arm, which provides a finer tuning of the delay (in steps of 140 as) as a function of the applied voltage. The linear polarization of either beam, at $\omega$ and $2\omega$, respectively, is rotated independently by half-wave retarders (Thorlabs, WPH05M-1550 and WPH05M-808) inserted into each arm of the interferometric stage. Another short-pass dichroic mirror (Thorlabs, DMSP950) recombines the two pumps, which are then focused onto the back-focal plane (BFP) of the objective (Nikon, CFI Plan Fluor 60XC, NA 0.85) by means of an achromatic lens doublet (Thorlabs, AC254-500-B), resulting in an almost collimated beam on the sample plane, with a spot size of about 15 μm. In this way, a large portion of the 100×100μm area metasurface is excited. The generated nonlinear signal is collected by the same objective in epi configuration and separated from the excitation by a long-pass dichroic mirror (Thorlabs, DMLP650). Here, a pair of achromatic doublets (Thorlabs, AC508-500-B) are placed in a 4$f$ configuration to relay the BFP image onto a back-illuminated Si CCD sensor (Andor, iKon-M DU934P-BV). Spectral filters (Thorlabs, FESH1000 + FESH0700 + FBH520-40) reject leakages of the excitation beams and select the emission centred at a wavelength of 517 nm (3$\omega$). A linear polarizer (Thorlabs, LPVISB100-MP2) mounted on a motorized goniometer (Thorlabs, PRM1Z8) is employed to select the linear polarization of the upconverted light at 3$\omega$.

To retrieve the intensity upconverted by the metasurface in the various diffraction orders, we evaluated the optical throughput of all the elements in the collection path as well as the efficiency of the CCD camera at 3$\omega$ ($\lambda =$ 517 nm). Therefore, we experimentally assessed the transmittance of the objective (~0.8) and the employed spectral filters (DMLP650, FESH1000, FESH700, FBH520-40) (~0.92) along with that of the polarizer (~0.5), lenses and the reflectance of silver mirrors (~0.5), which agree well with the manufacturer's specifications. Finally, the photon-to-count conversion of the CCD camera, considering both its sensitivity and quantum efficiency, was evaluated to be ~0.18. Based on the optical throughput, we estimated about 0.03 camera counts per photon emitted by the metasurface. To evaluate the power detected per pixel (*i.e.* the intensity) at 3$\omega$ displayed in Figures 2 and 4 of the main manuscript we multiplied the counts measured at the CCD camera by the energy of the photon at 3$\omega$ and divided by the above conversion factor.

**Acknowledgements**

The authors acknowledge financial support from the European Union's Horizon 2020 Research and Innovation program through the project 'METAFAST' (Grant Agreement No. 899673)